\documentclass[doublespacing]{elsart}
\usepackage[english]{babel}

\usepackage{epsfig}
\usepackage{graphics}

\usepackage{amssymb}

\begin{document}

\begin{frontmatter}

\title{Scattering of Antiprotons
by Nuclei (Atoms) in the Range of Low Energies $E\lesssim
10^5$\,eV. Mirror Reflection, Diffraction, and Channeling of
Antiprotons in Crystals.}

\author{V.G. Baryshevsky}

\address{Research Institute for Nuclear Problems, Belarusian State
University, 11~Bobruiskaya Str., Minsk 220030, Belarus}
\ead{bar@inp.bsu.by, v\_baryshevsky@yahoo.com}

\begin{abstract}
Studying antiproton scattering by nuclei (atoms) in the range of
low energies, we found out that  the increase in the
antiproton-nucleus scattering amplitude through Coulomb
 interaction provides the possibility to
investigate spin-dependent processes accompanying the interaction
of  antiprotons with nuclei (polarized or unpolarized) by means of
 mirror reflection of antiprotons from the vacuum-matter boundary,  diffraction and
channeling (surface diffraction and channeling) of antiprotons in
crystals.
\end{abstract}

\end{frontmatter}
\section{Introduction}

The progress in development of the Facility for Low-Energy
Antiproton and Ion Research (FLAIR) has spurred the rapid
development of low-energy antiproton physics \cite{1,2}. In
particular,  it was shown in \cite{PHL} that Coulomb scattering
leads to an increase in the real part of the amplitude of elastic
scattering of antiprotons by nuclei as the energy of antiprotons
decreases. As a result,  the effective interaction energy between
antiprotons and matter also rises with decreasing antiproton
energy. This gives the possibility to observe antiproton spin
rotation in a nuclear pseudo-magnetic field in matter with
polarized nuclei. The observation of this phenomenon is not only
of general physical interest but can also give information about
the amplitude of coherent elastic zero-angle scattering in the
range of low antiproton energies.

In this paper we show  that the  increase in the
antiproton-nucleus scattering amplitude in the range of low
energies through Coulomb
 interaction  makes it possible to
investigate many spin-dependent processes that accompany the
interaction of  antiprotons with nuclei (polarized or unpolarized)
by means of
 mirror reflection of antiprotons from the vacuum-matter boundary,  diffraction and
channeling (surface diffraction and channeling \cite{HENO}) of
antiprotons in crystals.

\section{Mirror Reflection of Antiprotons from the Surface of
Matter}

Before we start to consider mirror reflection of antiprotons from
the surface formed by the vacuum-matter boundary, let us make some
general remarks about mirror reflection of particles.

    It is well known \cite{gur-tar,landau} that the mirror-reflection coefficient
    $R$ is defined by the refractive index $n$ of a wave in
    matter. For such particles as, say,  neutrons (for light and
    $\gamma$-quanta, whose polarization is orthogonal to the
    mirror-reflection plane), the mirror-reflection coefficient has the
    form \cite{gur-tar,landau}
    \begin{equation}
    \label{eq1}
    R=|F|^2,
    \end{equation}
 \begin{equation}
    \label{eq2}
    F= - \frac{\sqrt{n^2-\cos^2\varphi}-\sin \varphi}{\sqrt{n^2-\cos^2\varphi}+\sin
    \varphi},
    \end{equation}
    where $\varphi$ is the grazing angle (see Fig.1).
   \begin{figure}[htbp]
\label{fig1}
\begin{center}
     \resizebox{60mm}{!}
       {\includegraphics{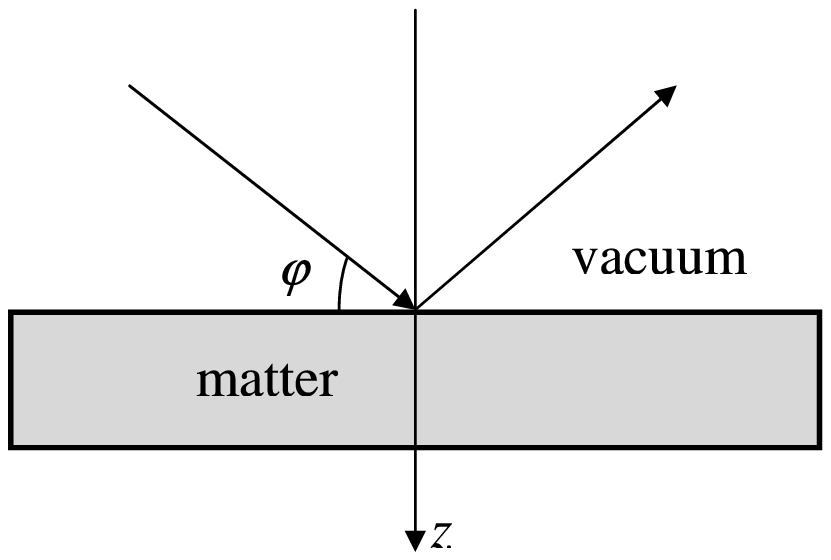}}\\
\caption{}
\end{center}
\end{figure}

    The refractive index (of particles or $\gamma$-quanta) can be
    expressed as follows \cite{7,8}:
    \begin{equation}
    \label{eq3}
    n^2=1+\frac{4\pi\rho}{k^2}f(0),
    \end{equation}
    where $\rho$ is the number of scatterers (nuclei, atoms) per
    cubic centimeter of matter and $f(0)$ is the amplitude of
    zero-angle coherent elastic scattering.
    According to (\ref{eq1})--(\ref{eq3}), the mirror-reflection
    coefficient is determined by the zero-angle scattering
    amplitude. But at the same time, the wave scattered in the
    mirror-reflection direction obviously makes a certain angle $\vartheta$
 relative to the incident direction. Thus, the  amplitude of this wave is determined by the superposition of waves  scattered
 at the angle $\vartheta$, i.e.,  instead of
 $f(0)$, the amplitude  $F$ is now
 determined by the amplitude  $f(\vartheta)$ of coherent elastic
 scattering at the angle~$\vartheta$.

 Indeed, let us consider elastic scattering of a wave
 $e^{i\vec k\vec r} = e^{i\vec k_{\perp}\,\vec r_{\perp}}\,e^{ik_z z}$
by a set of scatterers located in the plane $z=z_0$.  Here $\vec
k_{\perp}$ is the wave vector's component perpendicular to the
$z$-axis (parallel to the plane where the scatterers are placed),
$k_z$ is the component of the particle's wave vector
 that is parallel to the $z$-axis. Upon summation of the
spherical waves produced by the scatterers  in the plane $z=z_0$,
we obtain the following expression for the amplitude of a
mirror-reflected wave $F_1$ \cite{HENO}:
\begin{equation}
\label{eq4} F_1 =\frac{2\pi i \rho'}{k_z}f(\vec k'- \vec k)e^{2 i
k_z z_0},
\end{equation}
where $|\vec k'| = |\vec k|$, $\vec k'$ is the wave vector  of the
scattered particle,  which has the components $\vec k_{\perp}' =
\vec k_{\perp}'$ and $k'_z=-k_z$, and $\rho'$ is the density of
scatterers in the considered plane (the number of scatters in
cm$^2$ of the plane).

If the layer $[0,z]$ contains $m$ number of planes, then the
amplitude of the wave reflected by these planes  can be written in
the form
\begin{equation}
\label{eq5}
 F=\frac{2\pi i \rho'}{k_z} f(\vec k' - \vec k)\sum_{m}
e^{2i k_z z_m},
\end{equation}

Passing to a continuous distribution of planes in the layer $[0,
z]$, i.e., replacing the summation in (\ref{eq5}) by integration,
we finally obtain \cite{HENO}
\begin{equation}
\label{eq6} F = - \frac{\pi\rho}{k_z^2}f(\vec k' - \vec k)= -
\frac{4\pi\rho}{|\vec k'-\vec k|^2} f(\vec k '-\vec k).
\end{equation}

According to (\ref{eq6}), microscopic summation of waves scattered
at the vacuum-matter boundary yields the formation of a wave
reflected in the direction determined by the laws of classical
optics. Against (\ref{eq1}), its amplitude is defined not by the
amplitude $f(0)$, but by the amplitude of scattering at a nonzero
angle $\vartheta$ equal to a double grazing angle $\varphi$, i.e.,
$\vartheta = 2\varphi$ \cite{HENO}. Equation (\ref{eq1}) is valid
for thermal and ultracold neutrons because of prevailing isotropic
S-scattering by nuclei ($f(\vartheta)= f(0)$). It is also valid
for photons, because at wavelengths much greater than the size of
the atom, dipole scattering, which is also isotropic under the
considered polarization, occurs.

According to (\ref{eq5}), the  amplitude $F$ of mirror reflection
rises with decreasing  $k_z=k \sin \varphi$, i.e., with decreasing
grazing angle $\varphi$. When $F$ reaches the values close to
unity, equation (\ref{eq6}) for $F$ is no longer valid. Mirror
reflection for this range of angles is considered, e.g., in
\cite{HENO}.

In view of (\ref{eq6}), the amplitude $F$ is close to unity in the
range where the grazing angles $\varphi$ satisfy the condition
\begin{equation}
\label{eq7} \sin^2 \varphi \sim \frac{\pi\rho |f|}{k^2}.
\end{equation}
If in estimating the angle $\varphi$ we use a typical value of the
amplitude of scattering via nuclear interaction, $|f|\sim
10^{-12}$\,cm,   then (\ref{eq7}) readily yields that for
antiproton energies from 100 to 1000 eV, the value of $F$ can be
close to unity only when the grazing angles are very small:
$\varphi \leq 10^{-4}\ \div 10^{-5}$. Further in our consideration
we shall demonstrate that owing to the interference of Coulomb and
nuclear interactions and the increase in the amplitude of nuclear
scattering of antiprotons with decreasing energy \cite{PHL}, the
coefficient of mirror reflection for antiprotons becomes
noticeable even at much larger grazing angles $\varphi$, making it
possible to use the phenomenon of mirror reflection to investigate
scattering of slow antiprotons by nuclei.

According to (\ref{eq6}), the mirror-reflection coefficient $R$
can be written as follows:
\begin{equation}
\label{eq8} R=|F|^2=\left|\frac{\pi\rho}{k_z^2} f(\vec k'-\vec
k)\right|^2=\frac{\pi^2\rho^2}{k^4\sin^4\varphi} |f(\vec k'-\vec
k)|^2.
\end{equation}

There are two interactions responsible for antiproton scattering
by nuclei (atoms): Coulomb and nuclear (here we neglect the
magnetic interaction of antiproton and electron spins).
Consequently, the scattering amplitude $f$ can be presented as a
sum of two amplitudes:
\begin{equation}
\label{eq8} f= f_{Coul} + f_{N},
\end{equation}
where $ f_{Coul}$ is the amplitude of a purely Coulomb scattering
and $f_{N}$ is the amplitude related to nuclear interaction (it
contains the contribution from Coulomb interaction that affects
nuclear scattering \cite{PHL}). Let us note that the contribution
to the formation of a mirror-reflected wave comes from  elastic
scattering, in which the state of the target does not change. As
is known, the scattering amplitude in this case can be presented
as a product of the amplitude of elastic scattering by an
infinite-mass nucleus (atom)(the reduced mass equals the mass of
the incident particle) into the Debye-Waller factor $e^{-w(\vec k'
-\vec k)}$ describing the effect produced by thermal oscillations
of nuclei (atoms) in matter  on the process of scattering
\cite{gur-tar}. In subsequent consideration, by the amplitude $f$
we shall  mean  scattering by an infinite-mass nucleus. As a
result, we have ($\varphi\ll 1$)
\begin{equation}
\label{eq9} R=\frac{\pi^2\rho^2}{k^4
\varphi^4}\left(|f_{Coul}(\vec k'-\vec k)|^2 + 2 \texttt{Re}
f_{Coul}(\vec k'-\vec k) f_{N}^* + |f_{N}|^2\right) e^{-2w(\vec k'
- \vec k)}.
\end{equation}

For further consideration we need to compare the amplitudes of
Coulomb and nuclear elastic scattering of antiprotons by nuclei
(atoms).

\section{The Amplitude of Antiproton Scattering by a Nucleus (Atom)
at Low Temperatures}

The scattering amplitude relates to the $T$-matrix as \cite{7,8}
\begin{equation} \label{eq10} f_{ba}= -
\frac{m}{2\pi\hbar^2}\langle\Phi_{b}|T|\Phi_{a}\rangle,
\end{equation}
where $|\Phi_{a(b)}\rangle$ is the wave function describing the
initial (final) state of the system "incident particle--atom
(nucleus)". The wave functions $|\Phi_a\rangle$  are the
eigenfunctions of the Hamiltonian
${H}_0=H_p(\vec{r}_p)+H_A(\vec{\xi}\,, \vec{r}_{\mathrm{nuc}})$,
i.e., {${{H}_0|\Phi_a\rangle=E_a|\Phi_a\rangle}$};
$H_p(\vec{r}_p)$ is the Hamiltonian of the particle incident onto
the target; $\vec{r}_p$ is the particle coordinate;
$H_A(\vec{\xi}\,, \vec{r}_{\mathrm{nuc}})$ is the atomic (nuclear)
Hamiltonian; $\vec{\xi}$ is the set of coordinates of the atomic
electron; $\vec{r}_{\mathrm{nuc}}$ is the set of coordinates
describing the atomic nuclei.

The Hamiltonian $H$ describing the particle--nucleus  interaction
can be written as:
\begin{equation}
\label{a3} H=H_0+V_{\mathrm{Coul}} (\vec{r}_p\,,\vec{\xi}\,,
\vec{r}_{\mathrm{nuc}})+V_{\mathrm{nuc}}(\vec{r}_p\,,
\vec{r}_{\mathrm{nuc}})\,,
\end{equation}
where $V_{\mathrm{Coul}}$ is the energy of  Coulomb interaction
between the particle and the atom, and $V_{\mathrm{nuc}}$ is the
energy of nuclear interaction between the particle and the atomic
nucleus.

According to the quantum theory of reactions \cite{7,8}, in the
case of two interactions, the matrix element of the operator $T$,
which  describes the system's transition from the initial state
$|\Phi_a\rangle$ into the final state $|\Phi_b\rangle$, can be
presented as a sum of two terms:
\begin{equation}
\label{a9} T_{ba}=T_{ba}^{\mathrm{Coul}}+T_{ba}^{\mathrm{N}}=
\langle\Phi_b|T_{\mathrm{Coul}}|\Phi_a\rangle+\langle\varphi_b^{(-)}|T_{\mathrm{N}}|\varphi_a^{(+)}\rangle\,,
\end{equation}
where the first term, $T_{ba}^{\mathrm{Coul}}$, describes the
contribution to the T-matrix that comes from the Coulomb
scattering alone, the operator
\begin{equation}
\label{a10}
T_{\mathrm{Coul}}=V_{\mathrm{Coul}}+V_{\mathrm{Coul}}(E_a-H_0+i\varepsilon)^{-1}T_{\mathrm{Coul}}\,,
\end{equation}
and the second term describes the contribution to the T-matrix
that comes from nuclear scattering and accounts for the distortion
of waves incident onto the nucleus, $\varphi^{(\pm)}$, which is
caused by the Coulomb interaction. The operator
\begin{eqnarray}
\label{a11}
T_{\mathrm{N}}&=&V_{\mathrm{nuc}}+V_{\mathrm{nuc}}(E_a-H_0- V_{\mathrm{Coul}}+i\varepsilon)^{-1}T_{\mathrm{N}}\nonumber\\
&=& V_{\mathrm{nuc}}+V_{\mathrm{nuc}}(E_a-H_0-
V_{\mathrm{Coul}}-V_{\mathrm{nuc}}+i\varepsilon)^{-1}V_{\mathrm{nuc}}\,,
\end{eqnarray}
and  the wave functions $\varphi_{a}^{(\pm)}$ describe the
interaction between particles and atoms via the Coulomb
interaction alone ($V_{\mathrm{nuc}}=0$) \cite{7,8,9}:
\begin{equation}
\label{a6} \varphi_{a}^{(\pm)}=\Phi_a+(E_a-H_0\pm
i\varepsilon)^{-1}V_{\mathrm{Coul}}\, \varphi_a^{(\pm)}\,,
\end{equation}
the wave function $\varphi_a^{(+)}$ at large distances has the
asymtotics of a diverging spherical wave, and the wave function
$\varphi_a^{(-)}$ at large distances has the asymtotics of a
converging spherical wave \cite{7,8,9}.

Let us give a more detailed consideration of the matrix element
$\langle\varphi_b^{(-)}|T_{\mathrm{N}}|\varphi_a^{(+)}\rangle$.
Because nuclear forces are short--range, the radius  of the domain
of  integration in this matrix element is of the order of the
nuclear radius (of the order of the radius of action of nuclear
forces in the case of the proton). The Coulomb interaction,
$V_{\mathrm{Coul}}$, in this domain is noticeably smaller than the
energy of nuclear interaction, $V_{\mathrm{nuc}}$. We can
therefore neglect the Coulomb energy in the first approximation in
the denominator of (\ref{a11}), as compared to $V_{\mathrm{nuc}}$.

As a result, the operator $T_{\mathrm{N}}$ is reduced to the
operator describing a purely nuclear interaction between the
incident particle and the nucleus. The effect of Coulomb forces on
nuclear interaction is described by wave functions
$\varphi_{ba}^{(\pm)}$  (distorted--wave approximation \cite{8}).

In the range of antiproton energies of hundreds of
kiloelectronvolts and less, the de Broglie wavelength for
antiprotons is larger than the nuclear radius. Therefore, in
(\ref{a9}) for $T_{ba}^{\mathrm{N}}$, one can remove the wave
functions $\varphi_{a(b)}^{\pm)}$ outside the sign of integration
over the coordinate of the antiproton center of mass, $\vec{R}_p$,
at the location point of the nuclear center of mass,
$\vec{R}_{\mathrm{nuc}}$. As a result, one may write the following
relationship \cite{PHL}:
\begin{equation}
\label{a12} T_{ba}^{\mathrm{N}}=g_{ba}T_{ba}^{\mathrm{nuc}}=
\langle\varphi_b^{(-)}(\vec{R}_p=\vec{R}_{\mathrm{nuc}})|\varphi_a^{(+)}
(\vec{R}_p=\vec{R}_{\mathrm{nuc}})\rangle T_{ba}^{\mathrm{nuc}}\,,
\end{equation}
where $T_{ba}^{\mathrm{nuc}}$ is the matrix element describing a
purely nuclear interaction (in the absence of Coulomb interaction)
between the incident particle and the nucleus. The factor
$g_{ba}=\langle\varphi_b^{(-)}(\vec{R}_p=\vec{R}_{\mathrm{nuc}})|\varphi_a^{(+)}
(\vec{R}_p=\vec{R}_{\mathrm{nuc}})\rangle$ appearing in {Eq.~
(\ref{a12})} defines the probability to find the antiproton (the
negative hyperon, e.g. $\Omega^{-}\, , \Sigma^{-}$) at the
location of the nucleus.

{From} (\ref{eq10}), one can derive the following expression for
scattering amplitude:
\begin{equation}
\label{eq11} f_{ba}^N= - \frac{m}{2\pi\hbar} g_{ba}
T_{ba}^{nuc}=g_{ba}f _{ba}^{nuc},
\end{equation}
where $f _{ba}^{nuc}$ is the amplitude of particle scattering by
the nucleus in the absence of Coulomb interaction.

Thus the Coulomb interaction leads to a change in the value of the
amplitude of antiproton-nucleus scattering. Let us estimate the
magnitude of this change.

%%%%%%%%%%%%%%%%%%%%%%%%

In what follows we shall be primarily concerned with elastic
scattering. In this case $|\vec k| = |\vec k'|$, and it follows
that from the expression given in \cite{9} for the wave functions
$\varphi_b^{(-)}$ and $\varphi_a^{(+)}$, which describe particle
scattering in the Coulomb field, we can derive  the below
relationships for $g_{ba}$:
%%%%%%%%%%%%%%%%%%%%%%%%%%%%%%%%%

\begin{itemize}
\item for the case of repulsion, i.e., elastic scattering of
similarly charged particles
\begin{equation}
\label{a14}
g_{ba}^{\mathrm{rep}}=\frac{2\pi}{\kappa(e^{\frac{2\pi}{\kappa}}-1)}\,,\qquad
\kappa=\frac{v}{Z\alpha c}\,,
\end{equation}
where $v$ is the particle velocity, $Z$ is the charge of the
nucleus, $\alpha$ is the fine structure constant, $c$ is the speed
of light; \item for the case of attraction
\begin{equation}
\label{a15}
g_{ba}^{\mathrm{att}}=\frac{2\pi}{\kappa(1-e^{-\frac{2\pi}{\kappa}})}\,.
\end{equation}
\end{itemize}

With decreasing particle energy (velocity), $\kappa$ diminishes,
and for such values of $\kappa$ when $\frac{2\pi}{\kappa}\geq1$,
one can write
\begin{equation}
\label{a16} g_{ba}^{\mathrm{rep}}=\frac{2\pi\alpha Z
c}{v}\,e^{-\frac{2\pi\alpha Z c}{v}}\,,\qquad
g_{ba}^{\mathrm{att}}=\frac{2\pi\alpha Z c}{v}\,.
\end{equation}

According to (\ref{eq11}), with decreasing energy of positively
charged particles, the amplitude $f_{ba}^N$ diminishes rapidly
because of repulsion. For negatively charged particles, the
amplitude grows with decreasing particle energy (velocity).

These results for the amplitude $f^N_{ba}$ generalize a similar,
well-known  relationship for taking account of the Coulomb
interaction effect on the cross section of inelastic processes,
$\sigma_r$, \cite{9}.

So in the range of low energies, the amplitude $f(\vec k' -\vec
k)$ of antiproton (negative hyperon) scattering by a nucleus can
be presented in the form (for antiproton scattering in
ferromagnets, the magnetic interaction of antiprotons with
electrons in atoms should also be considered):
%штрих
\begin{equation}
\label{a18} f(\vec k' - \vec k)=f_{Coul}(\vec k' - \vec k) +
f_N(\vec k' - \vec k),
\end{equation}
where
\begin{equation} \label{a18b}
f_N(\vec k' - \vec
k)=\frac{2\pi\alpha Z c}{v}f_{nuc}(\vec k'- \vec k).
\end{equation}

Using (\ref{a18b}) and the expression for $f_{Coul}$ in \cite{9},
we can obtain the following expression for the modulus of the
Coulomb (Rutherford) scattering amplitude in the range of small
scattering angles ($\vartheta \ll 1$, $\vartheta>\frac{1}{kR}$):
\begin{equation}
\label{eq12a} |f_{Coul}(\vec k' -\vec k)|  = \frac{2 Z e^2}{m v^2
\vartheta^2} = \frac{Z e^2}{2 m v^2 \varphi^2}
\end{equation}
and then write the ratio for these amplitudes as
\begin{equation}
\label{eq12} \frac{|f_{N}|}{|f_{Coul}|}= 4\pi k \varphi^2
|f_{nuc}|.
\end{equation}

As a result, using for the characteristic nuclear amplitude the
estimate {${|f_{nuc}|\approx 10^{-12}}$}\,cm,{ ${k\leq 10^{10}\div
10^{11}}$}, $\varphi \sim 10^{-1}$, we can estimate the ratio
$\frac{|f_{nuc}|}{|f_{Coul}|}$ as
{${\frac{|f_{nuc}|}{|f_{Coul}|}\leq 3\cdot 10^{-4}\div 10^{-2}}$}.

In view of the above estimate, we can recast the expression for
the coefficient of reflection as follows:
\begin{equation}
\label{eq13} R=R_{Coul}\left(1+2\frac{\texttt{Re} e^{i\beta}
f_{nuc}^*}{|f_{Coul}|}\right) e^{-2w(\vec k' - \vec k)},
\end{equation}
where
$R_{Coul}=\frac{\pi^2\rho^2}{k^4\varphi^4}\left|f_{Coul}(\vec k' -
\vec k )\right|^2$ is the coefficient of mirror reflection due to
a purely Coulomb interaction,
$f_{Coul}=\left|f_{Coul}\right|e^{i\beta}$, and the term in
(\ref{eq9}) that is proportional to
$\frac{\left|f_{nuc}\right|^2}{\left|f_{Coul}\right|}$ is dropped
for its smallness. Let us recall that (\ref{eq13}) holds true for
such values of $R_{Coul}$ that are much less than unity.

Thus the coefficient of mirror reflection contains two
contributions: one comes from a purely Coulomb interaction and the
other is due to the  Coulomb-nuclear interference. This makes it
possible to obtain data about the amplitude of nuclear scattering
of antiprotons by the nucleus  from the experiments investigating
angular and energy dependence of $R$ on the grazing angle and the
energy of antiprotons.

\section{Spin Polarization of Antiprotons Reflected from the
Vacuum-Matter Boundary}

As is well known, spin-orbit interaction  during scattering causes
the initially unpolarized particle beam to become polarized
\cite{9}. In this case, the polarization vector of particles
appears to be orthogonal to the scattering plane, i.e., the plane
containing vectors $\vec k$ and $\vec k'$. If  the particle beam
had a nonzero polarization vector before the interaction, then a
left-right asymmetry  in the intensity of particle scattering is
observed.
 For slow neutrons, the spin-orbit
interaction is caused by the interaction $V_{so}$ of the neutron
magnetic moment and the nuclear electric field \cite{landau1}
\begin{equation}
\label{eq14} V^{neut}_{so}= i \frac{\mu_n \hbar}{m
c}\vec{\sigma}[\vec E(\vec r)\vec\nabla_{\vec r}],
\end{equation}
where $\mu_n$ is the neutron magnetic moment,
$\vec{\sigma}=(\sigma_x, \, \sigma_y,\, \sigma_z)$ are the Pauli
spin matrices, $\vec E$ is the electric field  at  point $\vec r$
where the neutron is located, and $m$ is the neutron's mass. First
experiments to observe the effect of spin-orbit interaction on
neutron scattering were performed by C.G. Shull \cite{shull}. (For
further experiments see, e.g., \cite{HENO}).

The presence of a charge in antiprotons appreciably affects  the
dependence of spin-orbit interaction on their magnetic moment. The
energy of spin-orbit interaction of antiprotons with nuclei is
determined by the electric field and has the form \cite{landau1}
\begin{equation}
\label{eq15} V_{so}^{ap}=- \left(\mu'+\frac{e\hbar}{4m c}\right)
\left(\vec{\sigma}\left[\vec E\frac{\hat{p}}{m}\right]\right),
\end{equation}
where $\mu'$ is the anomalous part of the antiproton's magnetic
moment, $e=-|e|$ is the antiproton charge, $|e|$ is the electron
charge, and $\hat{p}=- i\hbar \vec{\nabla}$ is the momentum
operator of the antiproton.

The energy dependence of the amplitude of spin-orbit scattering
also changes noticeably through the interference of Coulomb and
spin-orbit interactions. The energy dependence of the contribution
coming from the antiproton-nucleus strong interaction to the
amplitude of spin-orbit scattering changes, too.

The expression describing the amplitude of spin-orbit scattering
in a general form reads:
\begin{equation}
\label{eq16} F_{so}= F_{so}\vec {\sigma}[\vec k'\times \vec
k]=F_{so}\vec{\sigma}[\vec q\times\vec k],
\end{equation}
where $\vec q= \vec k'-\vec k$  is the  momentum transfer. It
should be noted that the general form of (\ref{eq16}) is clear
even without calculations and follows from the symmetry
considerations, being valid for all spin particles.

 Thus, the amplitude of mirror reflection from matter with unpolarized nuclei is a sum of three
 amplitudes: the amplitude of Coulomb scattering (or the amplitude of magnetic scattering by  electrons in ferromagnets), the
 particle-spin-independent amplitude of nuclear scattering, and
 the amplitude of spin-orbit scattering. Hence, the
 coefficient of mirror reflection contains the contributions coming from
 these amplitudes and their interference. Let us
 consider the contribution coming  to the coefficient of mirror
 reflection from the interference of Coulomb and spin-orbit
 interactions. Obviously, it is proportional to $ \vec \sigma [\vec k'\times\vec
 k]$.

 Let us choose the quantization axis to be parallel to the
 reflecting plane (the plane containing vectors $\vec k'$ and $\vec
 k$). It follows from (\ref{eq9}) and (\ref{eq16}) that due to Coulomb-nuclear interference, the reflection coefficients
 $R_{\uparrow\uparrow}$ and $R_{\downarrow\uparrow}$ for antiprotons
 with  spins parallel and antiparallel  to the direction of the axial vector $[\vec k\times \vec
 k']$ will differ.   Let an unpolarized antiproton beam be incident on the
 target. Such beam is represented by a coherent sum  of two
 beams with spins parallel and antiparallel to $[\vec k\times \vec
 k']$. A mirror-reflected beam appears partially
 polarized, and the degree to which the beam is polarized is
 determined by the difference $R_{\uparrow\uparrow} - R_{\downarrow\uparrow}$
 of mirror-reflection coefficients:
 \begin{equation}
 \label{eq17}
 p=\frac{R_{\uparrow\uparrow}-
 R_{\downarrow\uparrow}}{R_{\uparrow\uparrow}+  R_{\uparrow\downarrow}}
\end{equation}
When estimating the magnitude of the effect, we should take into
account that in the Born approximation, the amplitude $F_{so}$ is
purely imaginary, while the amplitude $F_{Coul}$ is real, and so
it is important that the imaginary part of $F_{Coul}$ be
considered. It can be readily found,  since we know that in the
range of small scattering angles, $\texttt{Im} F_{Coul}$ can be
set equal to $\texttt{Im}F_{Coul}(0)$ ($\texttt{Im}F_{Coul}(0)$ is
the amplitude of Coulomb scattering by a nucleus (atom) at zero
angle).

According to the optical theorem,
$\texttt{Im}F_{Coul}(0)=\frac{k}{4\pi}\sigma_{tot}$. The total
cross section of scattering by a screened Coulomb potential,
$V_{Coul}=\frac{Z e^2}{r}e^{-\frac {r}{ R_A}}$, can be written as
\begin{equation}
\label{eq18}
 \sigma= 16\pi \left(\frac{m Z e^2 R^2_A}{\hbar^2}\right)^2
 \frac{1}{1+ \frac{8 m E R_A^2}{\hbar^2}},
 \end{equation}
where $E$ is the particle energy.

It follows from (\ref{eq18}) that for antiproton energies greater
than the characteristic energy $E_A =\frac{\hbar^2}{4 z m
R^2_A}\approx 10^{-2}$ e V, the scattering cross section
$\sigma\sim
\frac{1}{E}\sim \frac{1}{v}$. %%% Z upper or lower case
As a result, $ \texttt{Im} F_{Coul}\simeq \mbox{const}$. The
amplitude of spin-orbit interaction contains the term
$\frac{2\pi\alpha Z c}{v}$ [see (\ref{eq11}), (\ref{a18b})] that
increases the amplitude of spin-orbit scattering of antiprotons by
nuclei (atoms), making it grow as $\sim \frac{1}{v}$ in the range
of small grazing angles (the momentum transfer in this case  is
required to be $q
> \frac{1}{R_A}$). As a result,  the contribution of spin-orbit
scattering to the coefficient of mirror reflection can be
estimated as
\[
R_{so}\sim \frac{\pi\rho p^2}{k^4\varphi^4}|\texttt{Im}
f_{Coul}|^2\frac{f_so}{\texttt{Im}f_{Coul}}\approx 10^{-1}\div
10^{-2}.
\]
Consequently, this process can be used to obtain  polarized
antiprotons in this low energy range.

Let us note here that when a polarized antiproton beam falls on
the surface, the particles' polarization vector rotates about the
quantization axis, i.e., about the direction $[\vec k'\times \vec
k]$. The phenomenon described here also occurs during the
diffraction reflection of antiprotons from a crystal's surface and
is caused by the interference between the spin-orbit amplitude and
the real part of the Coulomb amplitude in the case when crystal's
cells lack the center of symmetry. (Compare with a similar
phenomenon for slow neutrons \cite{HENO}, in which case  the
effect occurs through the interference of the spin-independent
part of nuclear scattering and the amplitude of spin-orbit
scattering.)

Now, let us suppose that a particle beam is incident on the
boundary between vacuum and matter with polarized nuclei. The
elastic scattering amplitude $\hat f(\vec k' - \vec k)$ in this
case depends on spin orientations of the incident particle, $\vec
S$,  and the nucleus, $\vec J$, i.e., the scattering amplitude is
the operator in the spin space of particle and nucleus.
Investigating refraction and mirror reflection, we are interested
in coherent elastic scattering, in which the nuclear spin state
remains unchanged. The scattering amplitude $\hat f_N(\vec k' -
\vec k)$, describing such scattering, is obtained by averaging
the total amplitude $\hat T$ using the nuclear spin density matrix
$\hat{\rho}_J$: $\hat f(\vec k' - \vec k) = \mbox{Tr}\,
\hat{\rho}\hat{T}(\vec k' - \vec k)$. (The general expression for
the amplitude $\hat T $ of scattering of a particle with spin
$S=\frac {1}{2}$ by a nucleus  with spin $J=\frac{1}{2}$ is given,
e.g., in \cite{9}.)

 Consequently, in this case the contribution from nuclear scattering to the amplitude
of a mirror-reflected wave can be written in the form:
\begin{equation}
\label{eq19a} \hat F_{pol}=-\frac{\pi\rho}{k_z^2}\hat f_N (\vec
k'- \vec k)
\end{equation}

 As a result, the amplitude of a
mirror-reflected wave can be presented in the form
\begin{equation}
\label{eq19} \hat F(\vec k'- \vec k) = F_{Coul}(\vec k' - \vec
k)+\hat F_{so}(\vec k ' - \vec k) + \hat F_{pol}(\vec k'-\vec k),
\end{equation}
 where $F_{Coul}$ is the amplitude of the
mirror-reflected wave that is due to Coulomb interaction, $\hat
F_{so} (\vec k' - \vec k) $ is the amplitude of the reflected wave
that is due to antiproton-nucleus (or antiproton-atom) spin-orbit
interaction, and { ${\hat F_{pol}(\vec k ' - \vec k)}$} is the
amplitude of antiproton scattering by a polarized nucleus (except
for those terms in the amplitude $\hat F_{pol}$ that describe
spin-orbit interactions).

Using (\ref{eq19}), we can find the intensity and polarization of
reflected particles. For example, the intensity of reflected
particles is related to spin orientation of incident particles by
the expression of the form
\begin{equation}
\label{eq20} I_{ref}=I_0\,\mbox{Tr}\,F\rho_0 F^+ = I_0
\,\mbox{Tr}\,F^+F \rho_0,
\end{equation}
where $\rho_0$ is the spin density matrix of the incident beam and
$I_0$ is the beam's intensity.

The polarization vector $\vec p$ of mirror-reflected particles has
the form
\begin{equation}
\label{eq21} \vec p=\frac{1}{I_{ref}}\,\mbox{Tr}\, \rho_0
F^+\,\frac{\hat{\vec S}}{S} F,
\end{equation}
where $\hat{\vec S}$ is the spin operator of particles; the spin
of antiprotons equals $1/2$, hence $\hat{\vec S}=\frac{1}{2}\vec
\sigma$.

It follows from (\ref{eq19}), (\ref{eq20}), and (\ref{eq21}) that
$I_{ref}$ and $\vec p$ depend on the interference of  the nuclear
amplitude $\hat F_{pol}(\vec k' - \vec k)$ and the Coulomb and
spin-orbit amplitudes.

The amplitude $\hat f_{N}$ that determines the reflection
amplitude $\hat F_{pol}$ has quite a complicated structure. For
slow antiprotons scattered at the angle $\vartheta\ll 1$ (the
grazing angle $\varphi\ll 1$), the amplitude $ \hat f_{N}$
coincides with a zero-angle scattering amplitude and has the form
\begin{equation}
\label{eq22} \hat f_{N}= A_0+ A_1 (\vec S\, \vec p_{t}) + A_2
(\vec S\,\vec e) (\vec e \,\vec p_t),
\end{equation}
where $\vec p_t$ is the polarization vector of the target. Using
(\ref{eq20}), (\ref{eq21}), and (\ref{eq22}), we can find the
intensity and polarization of reflected particles for each
particular case. These expressions yield that when unpolarized
antiprotons are incident on a polarized target, the reflected
antiprotons appear to be polarized. If the incident antiproton
beam is polarized, then the spin of the mirror-reflected beam
undergoes rotation (the rotation angle is estimated at the order
of $10^{-1}\div 10^{-2}$), and the intensity of the reflected beam
depends on the mutual orientation between the spin of incident
particles and the target polarization. The degree of polarization
that the initially unpolarized beam acquires through reflection
has the order of magnitude about $10^{-1}\div 10^{-2}$.

\section{Conclusion}

The analysis performed in this paper shows that the effects
described here can be used to obtain polarized beams of low-energy
antiprotons from unpolarized beams of low-energy antiprotons and
to study the polarization thereof.
%%%%%%%%%
By investigating the magnitudes of arising polarization and the
angle of spin rotation during mirror reflection, diffraction, or
channeling (surface diffraction and channeling),  we can
experimentally  measure the contributions coming from the
amplitudes $A_1$ and $A_2$.

\end{document}